\documentclass{lhep}

\def\be{\begin{equation}}
\def\ee{\end{equation}}
\def\bea{\begin{eqnarray}}
\def\eea{\end{eqnarray}}
\def\met{\not{\!{\rm E}}_T}
\def\zp{Z^\prime}

\usepackage[title]{appendix}
\usepackage{epsfig,latexsym}
\usepackage{hyperref}
\usepackage{amsmath}
\usepackage{verbatim}
\usepackage{color}
\usepackage{mathrsfs}
\usepackage{slashed}
\usepackage{amssymb}

\DeclareMathAlphabet{\mathpzc}{OT1}{pzc}{m}{it}

 \csname
@addtoreset\endcsname{equation}{section}

\def\gz0{\gamma^{0}}

\def\a{\alpha}

\def\x{\xi}

\def\vf{\varphi}

\def\cA{{\cal A}}
\def\cB{{\cal B}}

\def\beq{\begin{equation}}
\def\eeq{\end{equation}}
\def\bea{\begin{eqnarray}}
\def\eea{\end{eqnarray}}
\def\ba{\begin{array}}
\def\ea{\end{array}}
\def\bec{\begin{center}}
\def\ec{\end{center}}
\def\ba{\begin{align}}
\def\ena{\end{align}}

\def\12{\frac{1}{2}}

\def\comma{\,,\,}

\begin{document}

\title{String (In)Stability Issues with Broken Supersymmetry}

\author{J.~Mourad,\auno{1} A.~Sagnotti,\auno{2}}
\address{$^1$\sl APC, UMR 7164-CNRS, Universit\'e de Paris \\
10 rue Alice Domon et L\'eonie Duquet \\75205 Paris Cedex 13 \ FRANCE}
\address{$^2$\sl Scuola Normale Superiore and INFN\\
Piazza dei Cavalieri, 7\\ 56126 Pisa \ ITALY}

\begin{abstract}
We review the main results of our investigations motivated by the tadpole potentials of ten--dimensional strings with broken supersymmetry. While these are at best partial indications, it is hard to resist the feeling that they do capture some lessons of String Theory. For example, these very tadpole potentials lead to weak-string-coupling cosmologies that appear to provide clues on the onset of the inflation from an initial fast roll. The transition, if accessible to us, would offer a natural explanation for the lack of power manifested by the CMB at large angular scales. In addition, the same tadpole potentials can drive spontaneous compactifications to lower--dimensional Minkowski spaces at corresponding length scales. Furthermore, the cosmological solutions exhibit an intriguing ``instability of isotropy'' that, if taken at face value, would point to an accidental origin of compactification. Finally, symmetric static $AdS \times S$ solutions driven by the tadpole potentials also exist, but they are unstable due to mixings induced by their internal fluxes. On the other hand, the original Dudas--Mourad solution is perturbatively stable, and we have gathered some detailed evidence that instabilities induced by internal fluxes can be held under control in a similar class of weak--coupling type--$IIB$ compactifications to Minkowski space.
\vskip 12pt
\begin{center}
{\emph {Invited contribution to the special issue of Letters in High Energy Physics (LHEP) on \\ ``Swampland and String Theory Landscape'', edited by I.~Antoniadis, K.~Benakli and E.~Dudas}}
\end{center}
\end{abstract}

\maketitle


\section{Introduction}

\begin{figure}[ht]
\centering
\includegraphics[width=50mm]{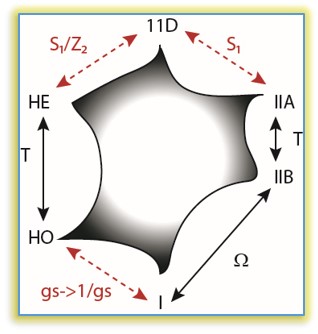}
\caption{\small The duality hexagon for ten--dimensional supersymmetric superstrings.}
\label{fig:susyduality}
\end{figure}

The highest achievement of the long and widespread effort devoted, over a few decades by now, to String Theory~\cite{strings}, is arguably the 10D--11D duality hexagon of fig.~\ref{fig:susyduality}. This links to one another the five ten--dimensional superstrings (of types $IIA$, $IIB$, $HE$, $HO$ and $I$), whose low--energy limits are captured by the three types of ten--dimensional supergravity~\cite{sugra,IIA, IIB, CP,gs}, thus granting some unprecedented clues that all of String Theory, despite the elusive nature of its foundations, stems somehow from a unique underlying principle. The first supergravity is the $IIA$ theory (whose fields are the \emph{non--chiral} combination $e_M^A$, $\psi_{M,L}$, $\psi_{M,R}$, $\lambda_L$, $\lambda_R$, $\phi$, $A_M$, $B_{MN}$ and $C_{MNP}$), where $\phi$ is the dilaton, the spinors are Majorana--Weyl and he subscript indicates their chirality. The second is the $IIB$ theory (whose fields are the \emph{chiral} combination $e_M^A$, $\psi_{M,L}$, $\lambda_R$, $\phi$, $a$, $B_{MN}$ and $D_{MNPQ}^+$), where $a$ is an axion, the spinors are Weyl, the two-form is complex and the four-form has a self--dual field strength. Finally, the third supergravity is the type-I theory, which rests on a reducible combination of the $N=1$ supergravity multiplet (whose fields are the \emph{chiral} combination $e_M^A$, $\psi_{M,L}$, $\lambda_R$, $\phi$, $B_{MN}$), where the spinors are again Majorana--Weyl, and super-Yang--Mills multiplets (with vector bosons $A_M^a$ and Majorana--Weyl spinors $\chi_L^a$). The nature of these multiplets grants the cancellation of gauge and gravitational anomalies via the Green--Schwarz mechanism~\cite{gs}, compatibly with their origin in String Theory. The heterotic $HO$ and $HE$ have thus $SO(32)$ and $E_8 \times E_8$ gauge groups~\cite{heterotic}, while only the former option is available for the type-I theory, due to the restriction in~\cite{cp}. Several people contributed to all this~\cite{tduality,orientifolds,pol,HW}, but Witten~\cite{witten1011} was arguably the driving force behind this monumental achievement, via a sapient combination of lessons drawn from String Theory and from the low--energy supergravity.

As is often the case, the highest achievements bring along the seeds of crisis, and this is strikingly true here. The five ten--dimensional superstrings are indeed connected, via the remaining side of the hexagon, to the eleven--dimensional supergravity of Cremmer, Julia and Scherk~\cite{CJS}, which was also, arguably, the highest point reached by supergravity in its development. While the additional links to this field theory were remarkable achievements, from a conceptual standpoint they led String Theory into a deep crisis, casting more than a doubt on the nature of its actual constituents. Surely enough, the dashed links in fig.~\ref{fig:susyduality} rest heavily, one way or another, on the existence of extended objects, branes, but these are, after all, generalizations of the solitons that had surfaced long before in Field Theory. On the other hand, the eleven--dimensional theory, which rests on an eleven--dimensional vielbein $e_M^A$, a Majorana gravitino $\Psi_M$ and a three-form gauge field ${\cal A}_{MNP}$, lacks the two typical signatures of strings: the dilaton $\phi$, which builds the string coupling $g_s = e^{\langle \phi \rangle}$, and a two-form gauge field. As a result, in this picture eleven--dimensional membranes~\cite{BST} replace somehow strings, since they couple naturally to a three--form potential, while the latter emerge merely from their wrapping around the eleventh dimension. These findings are usually summarized appealing to an unknown theory that will eventually encompass all different cases, different as they are, as special limits, but the very foundations of String Theory were thus invested by a high--intensity quake.
\begin{figure}[ht]
\centering
\includegraphics[width=80mm]{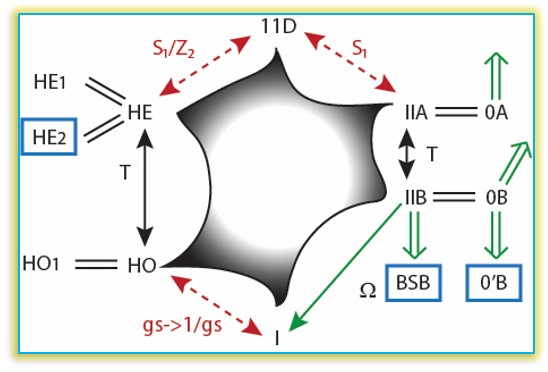}
\caption{\small The larger duality diagram including the ten--dimensional non--supersymmetric superstrings. The green lines identify orientifold projections, most of which were first considered in the \emph{Phys.~Lett.}~\textbf{B} paper with M.~Bianchi in~\cite{orientifolds}. The blue boxes identity the three non--tachyonic models: $HE2$ stands for the $SO(16) \times SO(16)$ model of~\cite{agmv}, $0'B$ for the $U(32)$ model of~\cite{as95} and BSB for Sugimoto's model in~\cite{bsb}.}
\label{fig:nonsusyduality}
\end{figure}

What is less widely appreciated is perhaps that the famous picture~in fig.~\ref{fig:susyduality} collects at most a fraction of the available options. If one insists on the world--sheet consistency rules of String Theory, there are in fact many more solutions in ten dimensions, which are indicated in fig.~\ref{fig:nonsusyduality}. The additional options~\cite{dh,sw,agmv,orientifolds,as95,bsb} lack space--time supersymmetry, and are thus a unique laboratory to gather some information on what String Theory tells us, at the most fundamental level, on the key issue of supersymmetry breaking. Surely enough, any attempt to connect String Theory with the Standard Model of Particle Physics is confronted with a bottom--up approach to supersymmetry breaking, but here String Theory itself is calling for an understanding of the phenomenon from a top--down perspective. The vast majority of the new options contain tachyons in their spectra, and following the fate of their vacua, while possible in principle, appears to date prohibitively difficult. For example, important progress was made in the early 2000's in connection with the tachyon of the open bosonic string~\cite{opentachyon}, but the corresponding closed--string tachyon is still fraught with mysteries. Hence, it appears reasonable to concentrate on the three new options, identified by blue boxes, where supersymmetry is broken, or is present but non--linearly realized, and yet the low--lying spectra contain no tachyons~\cite{ms_review}. There are two models of the first kind, the $SO(16)\times SO(16)$ heterotic string~\cite{dh,agmv}, whose massless spectrum contains states corresponding to $(e_M^A, B_{MN},\phi)$, together with adjoint vectors, left--handed fermions in the $(128,1)+(1,128)$ and right--handed fermions in the $(16,16)$, and the $U(32)$ $0'B$ orientifold~\cite{as95}, whose massless spectrum contains bosonic states corresponding to $(e_M^A,\phi,a,B_{MN},D_{MNPQ}^+)$, together with adjoint vectors and left--handed fermions in the $496$ +$\overline{496}$. The third string model is Sugimoto's $USp(32)$ string~\cite{bsb}, whose massless spectrum combines the states of type-I supergravity with massless vectors in the adjoint of $USp(32)$ and Majorana--Weyl fermions in the corresponding  (reducible) antisymmetric representation. The presence of a singlet spinor is no coincidence in this case: it is the goldstino that ought to be eaten by the gravitino in a spontaneous breaking of supersymmetry. Local supersymmetry in indeed present in this case, in a non--linear phase~\cite{dmnonlinear}, and yet one cannot even write a mass term for the gravitino, which is a Majorana--Weyl spinor--vector! Supersymmetry is broken by the simultaneous presence, in the vacuum, of branes and orientifolds preserving complementary portions of supersymmetry, and the non--dynamical nature of the latter grants the absence of tachyonic modes. We have often referred to this type of phenomenon as ``brane supersymmetry breaking'', after identifying its first manifestation in six dimensions in~\cite{bsb}, which resolved an old puzzle reviewed in~\cite{erice}. The lack of a mass term does not contradict any well-ascertained notion: all three models, and in this one in particular, are not defined around ten--dimensional Minkowski space, since the breaking of supersymmetry induces an important back-reaction. This is signalled by the emergence of a runaway (``tadpole'') potential for the dilaton, which takes the form
\beq{}
\Delta {\cal S} \ = \ - T \int \sqrt{-g} \,d^{10}x \, e^{\gamma\,\phi}  \label{tadpole_potential}
\eeq
in the Einstein frame.
Here $\gamma = \frac{3}{2}$  in the two orientifold models, the $0'B$ with gauge group $U(32)$ and Sugimoto's $USp(32)$ model that hosts, as we have seen, the intriguing phenomenon of ``brane supersymmetry breaking''. In both, the specific value of $\gamma$ reflects the origin of $T$ from the residual tension of D-branes and orientifolds. On the other hand  $\gamma = \frac{5}{2}$ for the $SO(16) \times SO(16)$ string, where the back-reaction first manifests itself in the torus amplitude. The reader is kindly asked to take notice of these specific values, since they will play a role in the ensuing discussion.

One should also meditate on another conundrum brought up by fig.~\ref{fig:nonsusyduality}. The very presence of a fundamental string model in ten dimensions where supersymmetry is non--linearly realized should be regarded, in our opinion, as a puzzle that adds to the surprising link to eleven dimensions, since after all we are used to think of non--linear realizations as limiting forms of linear ones that emerge in singular limits, when one or more members of a multiplet disappear somehow from the spectrum. Are we missing something here?

After motivating the interest in the three non--tachyonic models, if need there be, the real issue is the line of attack for addressing the key questions posed by them. The problem is both technical and conceptual in its nature: one has in mind the full--fledged String Theory, but the low--energy effective field theory is the only tool effectively at our disposal today. Extracting information from it about String Theory proper is not an easy task, since the two are connected by a double expansion, in powers of $g_s$ and in powers of the curvature in units set by the string scale $\frac{1}{\sqrt{\alpha'}}$. When both couplings are small, the low--energy effective field theory is expected to yield reliable indications but, as we are about to see, this ideal setting can be at best approached.

With this proviso, we shall now review the information that has been gathered, so far, on the behavior of the three ten--dimensional string models with broken supersymmetry and yet no tachyonic modes. It concerns the two distinct contexts
of cosmological solutions and static compactifications, to which we devote the next two sections.

\section{Cosmological Solutions} \label{ref:cosmo}

The study of cosmological solutions in the presence of the tadpole potential~\eqref{tadpole_potential} started with the exact solution
\bea{}
e^\phi \!\!&=&\!\! e^{\phi_0}\,\left|\alpha\, t^2\right|^\frac{1}{3}\,e^{ - \ \frac{3}{4}\, \alpha t^2} \ , \qquad\qquad \alpha \ \sim T \ , \label{dm_cosmo} \\
ds^2 \!\!&=&\!\! \left|{\alpha} \,t^2\right|^\frac{1}{18} e^{\, \alpha \,\frac{t^2}{8}}\  \eta_{\mu\nu}dx^\mu dx^\nu - e^{-\frac{3}{2}\, \phi_0}\,\left|{\alpha}\, t^2\right|^{-\frac{1}{2}} e^{\, \frac{9}{8}\, \alpha\, t^2} dt^2  , \nonumber
\eea{}
built in~\cite{dm_vacuum}, which was then generalized to arbitrary values of $\gamma$ in~\cite{russo}. The key lesson of this set of cosmologies was exposed in~\cite{dks}: it is the presence of a sharp transition between two types of behavior as $\gamma$ overcomes the \emph{critical value} $\gamma=\frac{3}{2}$. For brevity, here we have only displayed the simplest expressions for the three models of interest, which correspond to $\gamma=\frac{3}{2}$ and apply to the two orientifolds. One can see clearly that, as the variable $t$ grows, $e^\phi$ grows, reaches a finite maximum and then decreases, which justifies the pictorial name ``climbing scalar''. The range of values attained by the string coupling $g_s$ in these cosmologies is thus bounded from above.

\begin{figure}[ht]
\centering
\includegraphics[width=30mm]{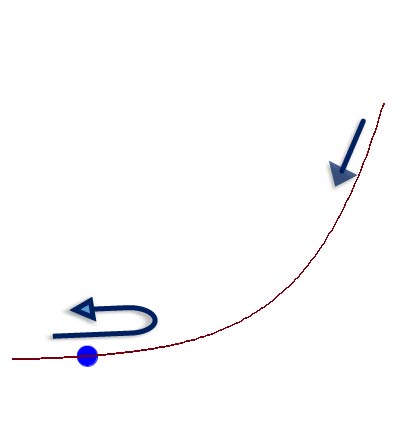}
\caption{\small The two widely distinct scenarios of a climbing and a descending scalar.}
\label{fig:climbing}
\end{figure}

Referring to fig.~\ref{fig:climbing}, we kindly ask the reader to pretend, initially, that the exponential be a horizontal line. For $\gamma=0$ the scalar field ought then to afford two distinct options: it should be able to proceed from large to small values, or alternatively from small to large ones, while loosing some of its energy due to cosmological friction. In the presence of a mild exponential potential, with a small positive value of $\gamma$, one could well associate to these two options the names of ``descending'' and ``climbing'' dynamics, although clearly in the latter case the scalar would climb up to a point to then revert its motion and start a descent. In String Theory, however, the two options entail a big difference: \emph{ the string coupling is unbounded for a ``descending'' scalar and is bounded for a ``climbing'' scalar}. The latter scenario can thus unfold entirely within the weak coupling region, providing information that is more reliable for String Theory as a whole.

The key result is that only the climbing behavior is possible for $\gamma \geq \frac{3}{2}$, and the transition point lies precisely at the value pertaining to the non--tachyonic $U(32)$ and $USp(32)$ ten--dimensional orientifolds! As a result, in all three ten--dimensional models of~\cite{dh,agmv,as95,bsb} only the climbing dynamics is possible. It is then natural to wonder whether this dynamics might have left some tangible signs in the Universe. This led to think about inflation, which is usually considered in its steady state, but could have received its initial impulse by this mechanism. Of course, these top--down considerations would need be complemented by other information, capable of producing a mild potential capable of slowing down the descent. Even a simple combination
\beq{}
V \ = \ T\, e^{\gamma\,\phi} \ + \ T'\,e^{\gamma'\,\phi} \ ,
\eeq{}
with a small enough $\gamma'$, which could be induced by other branes or otherwise, could grant a slow--roll phase~\cite{dkps} in the eventual descent. The same dynamics obtains if one combines the potential of eq.~\eqref{tadpole_potential} with other milder contributions, and for instance with the celebrated Starobinsky potential~\cite{starobinsky}: in all cases the fast--roll injection of inflation would induce a depression in the power spectrum of scalar (and tensor) perturbations~\footnote{The reversal of the scalar motion at the end of its ascent would leave another distinct signature in primordial power spectra, a small peak, which seems however beyond reach for CMB experiments.}. With a short--enough inflation these scenarios could account for the lack of power in the low--$\ell$ CMB angular power spectrum, and the mechanism also entails a definite prediction: in this scenario the tensor--to--scalar ratio would grow, even by one order of magnitude~\cite{dkps,gklmns}, in the transition region to the usual power--like behavior of~\cite{cm}, whose behavior can be captured by the simple formula
\beq{}
P(k) \ = P_0 \ \frac{k^3}{\left[k^2+ \ \Delta^2\right]^{2 \,-\,\frac{n_s}{2}}} \ .
\eeq{}

A further piece of the puzzle emerged from the (in)stability analysis of~\cite{bms}. It has to do with a surprise concerning tensor perturbations: while spatially varying tensor perturbations are stable like all scalar ones, in the sense that they decay in the course of the cosmological evolution, homogeneous tensor perturbations behave differently, and experience a logarithmic growth within the early ascent,
\begin{eqnarray}
h_{ij}'' &+& \frac{1}{\eta}\  h_{ij}'\ + \ {\bf k}^2\,h_{ij} \ = \ 0 \ , \nonumber \\
h_{ij} &\sim& A_{ij}\, J_0(k\eta) \ + \ B_{ij}\, Y_0(k \eta) \quad ( {\bf k} \neq 0) \ , \nonumber \\
h_{ij} &\sim& A_{ij} \ + \ {B_{ij}} \ \log\left(\frac{\eta}{\eta_0}\right) \qquad ( {\bf k} = 0) \ .
\eea
The growth for $\mathbf{k}=0$ signals an \emph{instability of isotropy}. In this respect, the compactification of extra dimensions needed to connect String Theory to Nature might have resulted from a mere accident!
\begin{figure}[ht]
\centering
\begin{tabular}{cc}
\includegraphics[width=36mm]{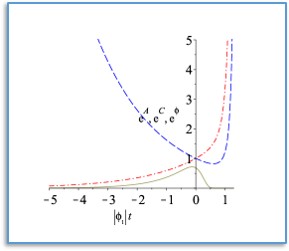} &
\includegraphics[width=38mm]{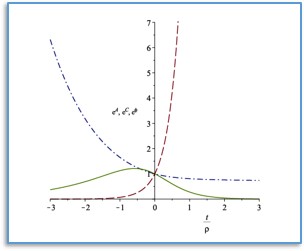} \\
\end{tabular}
\caption{\small $e^A$ (dashed), $e^C$ (dash--dotted) and $e^\phi$ (solid) for a ``critical'' anisotropic \emph{climbing scalar} cosmology with $\gamma=\gamma_c$, where the space--time $x$--coordinates undergo a bounce (left panel), and for an anisotropic \emph{climbing scalar} cosmology with $\gamma > \gamma_c$ where the internal directions undergo a contraction (right panel).}
\label{fig:new_cosmo}
\end{figure}

More recently, we have analyzed a wider class of solutions, which rest on metrics of the form
\beq{}
ds^2 \ =  \ - \ e^{2B(t)} dt^2 \ + \ e^{2 A(t)} dx^2 \ + \ e^{2 C(t)} dy^2 \ .
\eeq
These can describe extensions of the original setting of~\cite{dm_vacuum, russo}, or even further compactifications thereof, if the $y^i$ are identified periodically. Two interesting options that this adds to the preceding picture are displayed in fig.~\ref{fig:new_cosmo}. Concentrating on the more interesting case of climbing scalars, the cosmological expansion can be accompanied by bounces of the additional dimensions or, for large enough values of $\gamma$ (the critical value grows with the anisotropy and is larger than  $\frac{3}{2}$), or by additional dimensions undergoing a contraction. Therefore, additional toroidal dimensions have the option of compactifying dynamically.

A word of caution is needed before concluding this section. Truly enough, the climbing scenario grants a bounded string coupling, but high curvatures remain present close to the initial singularity. A systematic account of higher--derivative corrections might help with them, but the information gathered so far, which did solve some amusing puzzles with the climbing scenario, has not given positive indications in this respect~\cite{dc}.

\section{Static Solutions} \label{sec:static}

The lessons encoded in~\cite{dm_vacuum} are also a natural starting point to search for static compactifications in the presence of the tadpole term of eq.~\eqref{tadpole_potential}. There is a second surprise, which is however accompanied, as is usually the case, by corresponding difficulties. Let us begin by exposing the former. To this end, it will suffice to consider the analytic continuation of eq.~\eqref{dm_cosmo}, obtained letting $t \to i r$,
\bea{}
e^\phi \!\!&=&\!\! e^{\phi_0}\,\left|\alpha\, r^2\right|^\frac{1}{3}\,e^{ \, \frac{3}{4}\ \alpha r^2} \ , \qquad\qquad \alpha \ \sim T  \ ,  \label{dm_spatial} \\
ds^2 \!\!&=&\!\! \left|{\alpha} \,r^2\right|^\frac{1}{18} e^{\, -\,\alpha \,\frac{r^2}{8}}\  \eta_{\mu\nu}dx^\mu dx^\nu +  e^{-\frac{3}{2}\, \phi_0}\,\left|{\alpha}\, r^2\right|^{-\frac{1}{2}}\!\! e^{\,-\, \frac{9}{8}\, \alpha\, r^2} dr^2 \ . \nonumber
\eea{}
In this fashion the time--like $t$ becomes the spatial variable $r > 0$, and this simple operation brings along two striking effects that the reader can readily appreciate:
\begin{itemize}
    \item the string coupling becomes unbounded as $r\to \infty$;
    \item the length of the internal interval parametrized by $r$ becomes finite.
\end{itemize}

Remarkably, the tadpole potential~\eqref{tadpole_potential} drives a spontaneous compactification down to an interval of size proportional to $\frac{1}{\sqrt{T}}$! Notice that this setting is vastly different from the standard Kaluza--Klein circle reduction, where the size of the internal space would be a modulus. Yet, the theory maintains a genuine nine--dimensional interpretation, and gravity and gauge interactions survive in the resulting nine--dimensional spacetime with finite couplings, as shown in \cite{dm_vacuum}.

The curvature and the string coupling are both unbounded, however, so that there are two reasons to pause when attempting to ascribe these striking findings to String Theory proper, but there is a second surprise, which comes from the analysis of perturbations.

Briefly stated, perturbations can be classified, as is well known, according to the their behavior under the residual symmetry group, and can be studied accordingly. The analysis is aimed at determining the sign of $m^2$ for the available modes in the resulting Minkowski spacetime, and in this case tensor perturbations are simply stable. However, there are potentially dangerous scalar perturbations $A$, which affect the conformal--gauge metric according to
\beq{}
ds^2 \ = \ e^{2\Omega(z)} \left[ (1+ A) \,dx^\mu \, dx_\mu \ + \ (1-7A)\, dz^2 \right] \ .
\eeq
$A$ satisfies a complicated differential equation in $r$,
\bea{}
A'' &+& A'\left(24\, \Omega'\ - \ \frac{2}{\phi'} \ e^{2\Omega}\, V_{\phi} \right) \\  &+&
A\left(m^2 \ - \ \frac{7}{4} \ e^{2\Omega}\, V \ - \ 14 \, e^{2\Omega}\, \Omega'\, \frac{V_\phi}{\phi'} \right) \ = \ 0 \ , \nonumber
\eea{}
but one can make nonetheless definite statements on the sign of $m^2$, resorting to a very useful trick. This recurs, in various forms, in Physics and in Mathematics, and consists in recasting the equation in the form,
\bea{}
m^2\, \Psi &=& \left( b \ + \ {\cal A}^\dagger\, {\cal A} \right) \Psi \ , \label{AAdagpsi}
\eea{}
with suitable choices of ${\cal A}$ and ${\cal A}^\dagger$. In this case
\bea
{\cal A} &=& \frac{d}{dr} \ - \ \alpha(r) \ , \quad  {\cal A}^\dagger \ = \  - \ \frac{d}{dr} \ - \ \alpha(r) \ ,
\eea{}
where $\alpha$ is proportional to the coefficient of $A'$ above, and
\bea
b &=&  \frac{7}{2} \, e^{2\Omega}\, V \, \frac{1}{1 \ + \ \frac{9}{4}\, \alpha_O\, y^2}  \ > \ 0 \, . \label{AAdagger}
\eea
When applied in this context, these steps yield a definite prediction for the sign of $m^2$.
\emph{Up to a proper choice of boundary conditions}, the product ${\cal A}^\dagger\, {\cal A}$ is indeed a positive operator, and the stability of the model follows since $b$ turns out to be a positive function. In this fashion, one can conclude that the Dudas--Mourad vacuum for the orientifold models is \emph{perturbatively stable}, and similar steps lead to the same conclusion for its counterpart for the heterotic $SO(16) \times SO(16)$ string!

The preceding result was not at all guaranteed. It can be contrasted with the fate of another class of vacua, which were first considered in~\cite{gm} and were rediscovered and extended in~\cite{ms_16}. Let me focus again, for brevity, on the solution corresponding to $\gamma=\frac{3}{2}$, which is an $AdS_3 \times S^7$ vacuum described by a metric of the form
\beq{}
ds^{\,2}\ = \ R_{AdS}^2 \, \lambda_{\mu\nu}\, dx^\mu\, dx^\nu \ + \ R^2 \, \gamma_{ij}\, dy^i\, dy^j \ .
\eeq{}
Here the $H_3$-flux is required by the dilaton equation, which would select, in a similar fashion, an $H_7$-flux and thus an $AdS_7 \times S^3$ vacuum, for the heterotic $SO(16) \times SO(16)$ model. \emph{The crucial property here is that the $AdS$ and $S$ radii are proportional}. The background equations can be cast in the form
\bea{}
&& R_{AdS}^2\,V_0 \ = \ 12 \left( \sigma_3 \ - \ \frac{4}{3} \right) \ ,  \\
&& \sigma_3 \ = \frac{R_{AdS}^2}{2\,\beta}\, V_0^{'} \ =  \ 1 \ + \ 3\, \frac{R_{AdS}^2}{R^2} \ , \qquad \tau_3 \ = \ R_{AdS}^2 \, V_0^{''} \ , \nonumber
\eea{}
where $V_0$, $V_0'$ and $V_0''$ denote the values attained by the tadpole potential and its first two derivatives at the equilibrium value for $\phi$ determined by the flux.
We have kept the system in this form for a reason that will soon become apparent, although the actual values of the two parameters $\sigma_3$ and $\tau_3$ are easily derived from the complete set of equations, and are
\beq
\sigma_3 \ = \ \frac{3}{2} \ , \qquad \tau_3 \ = \ \frac{9}{2} \ . \label{sigmatau}
\eeq
Let us stress that this class of symmetric vacua would have, in principle, all it takes to expect that its indications be significant for the full--fledged String Theory: the flux can be adjusted at will, in such a way that curvature and string coupling be both small.

One can study again the behavior of perturbations in this setting, which is complicated by the mixing of different fields. However, the analysis boils down to tracking the squared masses of the different perturbations, without the need to address directly operator spectra, since they are explicitly determined by well--known properties of the Laplace operator on the internal sphere.
Everything works nicely for internal zero modes, but the Kaluza--Klein excitations bring along additional modes, which mix with the original ones complicating matters. Due to a well--known subtlety, which goes under the name of Breitenlohner--Freedman bounds~\cite{bf}, the transition to instability does not occur when $m^2$ changes sign, but when it falls below some well--defined negative values that depend on the nature of the fields involved. At any rate, the resulting eigenvalues reveal the emergence of unstable scalar modes. A similar behavior was found in~\cite{mns} for the $AdS_4$ vacuum of eleven--dimensional supergravity. We have not managed to eliminate the unstable modes by a sensible internal projection for the orientifolds, while one can simply do it for the heterotic solution invoking an $\mathbf{r} \to - \mathbf{r}$ identification in the internal space. This opens up another issue, related to the non--perturbative instabilities of these types of vacua, which were addressed in~\cite{pol_inst,ivano}. Still, it is amusing to note that the actual values of the two parameters $\sigma_3$ and $\tau_3$ in eq.~\eqref{sigmatau}, which identify the position of the cone in fig.~\ref{fig:AdSinstability}, lie somehow close to the border beyond which the compactification would be stable. One might wonder whether quantum corrections to the potential could be of help here.
\begin{figure}[ht]
\centering
\includegraphics[width=40mm]{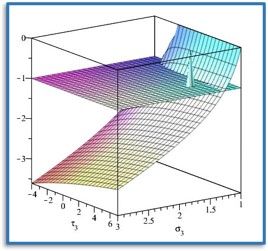}
\caption{\small \small A comparison between the lowest eigenvalue of $R_{AdS}^2\,{\cal M}^2$, which obtains for the internal angular momentum quantum number $\ell=3$, and the BF bound, which is -1 in this case. There are regions of stability for values of $\sigma_3$ close to 1, which correspond to $\frac{R^2}{R_{AdS}^2} > 9$ and negative values of $V_0$. . The small cone identifies the point corresponding to the actual tree--level values $\sigma_3=\frac{3}{2}$, $\tau_3=\frac{9}{2}$.}
\label{fig:AdSinstability}
\end{figure}

We can now review our recent work in~\cite{ms21_1}, which explores in some detail vacua described by metrics of the form
\beq{}
ds^2 \ = \ e^{2 A(r)} dx^2 \ + \ e^{2B(r)} dr^2 \ + \ e^{2 C(r)} dy^2 \ , \label{metric_space}
\eeq
while also allowing for internal fluxes respecting their symmetries, or the tadpole potential of eq.~\eqref{tadpole_potential}, or both. The main lessons that we thus gathered are the following:
\begin{itemize}
    \item the presence of internal fluxes can allow, by itself, solutions with $r$--intervals of finite length, and \emph{the string coupling can be bounded} in some of these examples, while the curvature remains singular;
    \item at one end of the interval, these solutions approach the behavior of corresponding solutions where supersymmetry is only partly broken, which are approached globally as the length of the interval tends to infinity;
    \item when the tadpole potential is present, even in combination with internal fluxes, a finite length of the $r$-interval becomes inevitable, since the effects of $T$ dominate  as $r\to \infty$, but the string coupling diverges in the asymptotic region. These systems are difficult to analyze in general, but we have managed to find an exact solution in the presence of $T$, for the value $\gamma=\frac{3}{2}$ that pertains to orientifolds, and also in the presence of an $H_7$ flux. The resulting equations read
   \begin{eqnarray}
 ds^2 \!\!\!&=&\!\!\! e^{\ - \  \frac{T\,e^{2\,x_2}\, r^2}{8\,\alpha'} \,-\,\frac{\chi_1\,r}{6}} \Bigg( \frac{dx \cdot dx}{\left[h\,\rho \, \sinh \left(\frac{r}{\rho}\right) \right]^\frac{1}{4}}  \nonumber \\ &+&   {\left[h\,\rho \, \sinh \left(\frac{r}{\rho}\right) \right]^\frac{3}{4}} \ dy \cdot dy \Bigg) \\
 \!\!\!&+&\!\!\! e^{\ - \  \frac{9\,T\,e^{2\,x_2}\, r^2}{8\,\alpha'} \,-\,\frac{3\left(\chi_1\,r\,+\,\chi_2\right)}{2}\,+\,2\,x_2} {\left[h\,\rho \, \sinh \left(\frac{r}{\rho}\right) \right]^\frac{3}{4}} \!\!dr^2 \ ,\nonumber\\
e^\phi \!\!&=&\!\! \frac{e^{\,\frac{3\,T\,e^{2\,x_2}\, r^2}{4\,\alpha'} \,+\,\chi_1\,r\,+\,\chi_2}}{\left[h\,\rho \, \sinh \left(\frac{r}{\rho}\right) \right]^\frac{1}{2}} \, ,
{\cal H}_{7} =  h\, \frac{\epsilon_{6}\, dr}{\left[h\,\rho \, \sinh \left(\frac{r}{\rho}\right) \right]^2} \ , \nonumber
\end{eqnarray}
    where $0 < r< \infty$, and clearly exhibit the inevitable dominance of the tadpole term for large values of $r$, together with a power--like behavior as $r\to 0$. As one could see taking a closer look at it, this behavior corresponds indeed to a supersymmetric limit.
\end{itemize}

In \cite{ms21_2} we have studied in detail a special type of compactification driven by an $H_5$ flux. It rests on the direct product of an internal torus and a finite internal interval, where the string coupling is constant, and is described by
\bea
&& ds^2 \,=\, \left[\frac{|H_5|}{\sqrt{2}}\,\rho\,\sinh\left(\frac{r}{\rho}\right)\right]^{\,-\,\frac{1}{2}} \!\! dx^2 \label{4d_inter_flux_H2} \\ &+& \left[\frac{|H_5|}{\sqrt{2}}\,\rho\,\sinh\left(\frac{r}{\rho}\right)\right]^{\,\frac{1}{2}}  \left( e^{\,-\,\frac{5\,r}{\rho\,\sqrt{10}}} \, dr^2 \, + \, e^{\,-\,\frac{r}{\rho\,\sqrt{10}}} \,d {y}^{\,2}\right)\, , \nonumber \\
&& {{\cal H}_5^{(0)}} \,=\,
\frac{H_5}{2} \left\{ \frac{dx^0 \wedge ...\wedge dx^3\wedge dr}{\left[\frac{|H_5|}{\sqrt{2}}\,\rho\,\sinh\left(\frac{r}{\rho}\right)\right]^2} \ + \ dy^1 \wedge ... \wedge dy^5\right\}  \, . \nonumber
\eea
Even in this case, one can show that the lower--dimensional Planck mass is finite, so that gravity remains at work in the resulting four--dimensional Minkowski spacetime.
One is thus improving on the Dudas--Mourad setting, albeit within a different context: these non--supersymmetric vacua concern a supersymmetric theory, the type--$IIB$ string, where the tadpole is absent and supersymmetry breaking is induced via the $H_5$ flux. As $\rho \to \infty$, one half of the ten--dimensional supersymmetry is recovered but the theory becomes five dimensional.

The detailed analysis of field perturbations is quite complicated, even after splitting them according to the residual symmetries of the background. Part of the complications originate from the unfamiliar fluctuations of the self--dual five-form field strength, which satisfies the first--order equation~\cite{IIB}
$ H \ = \ \star\, H$.

One can work conveniently in a radial gauge for the four--form potential, setting to zero all components $B_{rMNP}$ that point in the radial direction.
In order to convey a flavor of the nature of the problem, we can display the detailed form of the tensor equation,
\begin{eqnarray}
&&\!\!\!\!\! \partial_{[\mu}\,b^{(2)}{{}_{\nu]}}^{l m} + \frac{1}{2}\,\epsilon^{l m n p q}\,\partial_n\,b_{\mu \nu p q}   =  - \,  \frac{e^{-4A-4C}}{2}\, \epsilon_{\mu\nu\rho\sigma}\, \partial_r b^{\rho \sigma l m } \, ,\nonumber \\
&&\!\!\!\!\!  \partial_r\,{b^{(2)}}{{}_\mu}^{l m} =  {e^{2A+6C}}\left(\partial^{[l}\,{b_\mu}^{m]}\,+\,\frac{1}{2}\,{\epsilon^{\alpha\beta\gamma}}_\mu\,\partial_\alpha\,{b_{\beta\gamma}}^{lm} \right)\, ,
\nonumber \\
&&\!\!\!\!\!  \partial_\mu\,b^m \ - \ \partial_n\,{{b^{(2)}}_\mu}{}^{m n}  = e^{-2C}\left[\frac{H_5}{2} \ {h_\mu}^m  \,-\, {e^{-6A}}\, \partial_r\,{b_\mu}^m \right]\, ,
\nonumber \\
&&\!\!\!\!\!  \partial_r\,b^m   = e^{-2C}\left[\frac{H_5}{2}\,{h_r}^m\ - \ {e^{10C}}\left( \partial^m\,b\,-\,\partial_\mu\,b^{\mu m} \right)\right]\, ,
\\
&&\!\!\!\!\!  \partial_p\,b^p  =  \frac{H_5}{4}  \left( -  e^{-2A}\,{h_{\alpha}}^\alpha -  e^{-2B}\,h_{r r}  +  e^{-2C}\,{h_{i}}^i \right) +  e^{-8A} \,\partial_r\, b  \, , \nonumber
\end{eqnarray}
whose components look somewhat unfriendly even after the simplifications introduced by gauge fixing.
At any rate, making a long story short, all individual sectors of the spectrum can be related, via a sequence of field transformations, to Schr\"odinger--like equations as in eq.~\eqref{AAdagger}, but where the operators are generally matrix-valued.

The are no instabilities in the sectors carrying a vanishing momentum $\mathbf{k}$ in the internal torus, but as soon as one allows nonzero values of $\mathbf{k}$ new modes appear and, with them, mixings and potential instabilities. In this respect, the pattern is along the lines of what we saw in $AdS \times S$ vacua. We shall content ourselves, here, with a special example,
\begin{eqnarray}
&& M\,Z = m^2\, Z \ , \quad   \alpha =  \frac{C - A}{2} \ , \quad \beta = - \, \frac{5A+3C}{2} \label{extended_schrodinger} \\ \nonumber \\
M \!\!\!&=&\!\!\! \left(\begin{array}{cc} {\cal K}^2\,+ \left(-\,\partial + \alpha\right)_z\left(\partial + \alpha\right)_z  & \frac{k H_5}{\sqrt{2}}\, e^{2(A-3C)} \\ \!\!\!\!\!\!\!\!\!\!\frac{k H_5}{\sqrt{2}}\,e^{2(A-3C)} & \!\!\!\!\!\!\!\!\!\!\!\!\!\!{\cal K}^2 + \left(-\partial +\beta\right)_z\left(\partial + \beta\right)_z \end{array}\right) \nonumber \, ,
\end{eqnarray}
which can convey the general lesson. Notice that the diagonal terms, where ${\cal K}^2=k^2 e^{2(A-C)}$ and $k=\left|\mathbf{k}\right|$, contain manifestly positive contributions like those in eqs.~\eqref{AAdagpsi} and~\eqref{AAdagger}. The difficulty is due to the off--diagonal terms, which are linear in $k$ and can be associated to the $\sigma_1$ Pauli matrix. As such, they can add or subtract to the diagonal contributions, which are manifestly positive, and this is the origin of the tachyon problem. Notice also that, for large internal momenta $\mathbf{k}$, the diagonal terms dominate, so that the problem ought to concern a limited range of internal momenta.
\begin{figure}[ht]
\centering
\begin{tabular}{cc}
\includegraphics[width=28mm]{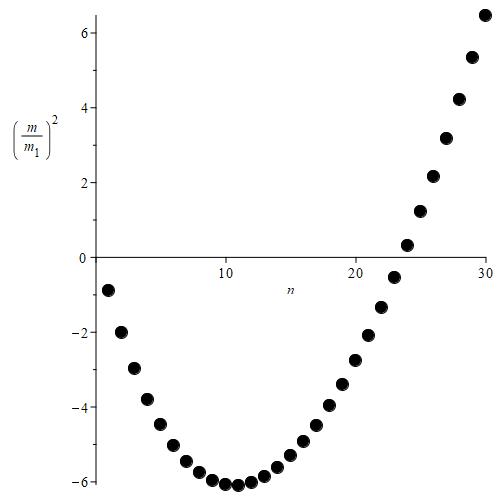}
&
\includegraphics[width=28mm]{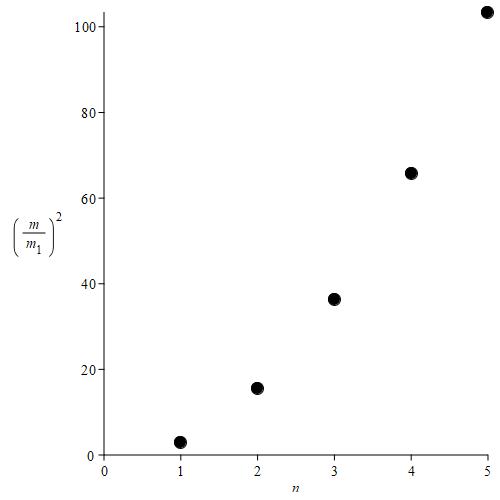} \\
\end{tabular}
\caption{\small Variational estimates of $m^2$ as a function of $\left|\mathbf{n}\right|$, where $\mathbf{k}=\frac{\mathbf{n}}{R}$ and $\mathbf{n} \neq 0$ in this sector, in units of $m_0^2=\left(\frac{\sqrt{2}}{H_5\,\rho^3}\right)$. The two figures refer to $\frac{R}{\rho}=10^{-1}$ (left panel) and to $\frac{R}{\rho}=2 \times 10^{-3}$ (right panel, where we have also compressed the vertical units by a factor 10). With the second choice all tachyonic modes disappear.}
\label{fig:m2_estimate}
\end{figure}

The actual effects of the off--diagonal terms depend crucially on the ratio between two scales, the length of the $r$-interval and the linear size, say $R$, of the internal torus, which \emph{can be chosen independently} in this setting, in sharp contrast to what we saw for $AdS \times S$ vacua. The low--energy analysis allows one to explore these regimes, insofar as both length scales lie well above the Planck and string scales. Consequently, \emph{there is some room to eliminate tachyonic modes here, if the torus size is sufficiently small compared to the size of the interval}.  This is illustrated in fig.~\ref{fig:m2_estimate}.

Let us conclude this section with a technical note. The Schr\"odinger--like systems at stake are very complicated, so that it is already amusing to see how their zero modes, if present, can be obtained in closed form. However, if one is looking for bounds on $m^2$, the evident analogies with non--relativistic Quantum Mechanics put at one's disposal what is perhaps its most powerful tool, the variational principle. Estimates of the ground--state energy for a Hermitian Schr\"odinger--like operator $\widetilde{M}$ obtained via a generic test function $\Psi$ lie above the actual ground--state energy,
\begin{eqnarray}
m_\Psi^2 \ = \ \frac{\langle \Psi | \widetilde{M} | \Psi \rangle}{\langle \Psi | \Psi \rangle} \ \geq \ m_0^2 \ ,
\end{eqnarray}
so that lower estimates are better ones.
The method can be amazingly accurate and lies at the heart of wide portions of Chemistry and Statistical Physics. There is first catch, however, in our case. All this is true provided the $\Psi$ functions satisfy proper boundary conditions at the ends of the $r$-interval. These were instrumental in selecting the test functions leading to the estimates in fig.~\ref{fig:m2_estimate}, and the setup to choose them is the subject of the next section.

\section{Boundary Conditions}

Let us consider a manifold ${\cal M}$ with a boundary $\partial{\cal M}$, while also referring directly to Fermi fields, whose formulation entails a richer structure. Our starting point is the variation of a matter action,
\beq
\delta\,{\cal S}_m \ = \ \int_{\cal M} d^{D}x \ e \ \left[ \delta {e_M}^A\, {{\cal T}^{M}}_A \ + \ \delta {\omega_M}^{AB} \ {{\cal Y}^M}_{AB} \right] \ ,
\eeq
which defines a (generally non symmetric) energy--momentum tensor ${{\cal T}^{M}}_A$ and an additional tensor ${{\cal Y}^M}_{AB} $, which will play a central role in the ensuing discussion. The vielbein will be covariantly constant, but a torsion tensor ${S^{P}}_{MN}$ will be generally present:
\begin{eqnarray}
&& D_M\,{e_N}^A \ \equiv \ \partial_M\,{e_N}^A \ + \ {{\omega}_M}^{AB}\ e_{NB} \ - \ {\Gamma^P}_{MN} \, {e_P}^A \ = \ 0 \ , \nonumber \\
&& {S^{P}}_{MN} \ = \ {\Gamma^P}_{MN} \ - \ {\Gamma^P}_{NM} \ .
 \nonumber
\end{eqnarray}
The invariance of the action under local Lorentz transformations, which act on vielbein and spin connection according to
\begin{eqnarray}
&&\delta\, {e_M}^A \ = \ \epsilon^{AB}\, {e_{MB}} \ , \qquad \delta\, {\omega_M}^{AB} \ = \ - \ D_M\,\epsilon^{AB} \ ,
\end{eqnarray}
yields a first Bianchi identity,
\begin{eqnarray}
&& D_M\, {{\cal Y}^M}_{AB} \ - \ {S^P}_{PM} \,{{\cal Y}^M}_{AB} = \frac{1}{2} \left( {{\cal T}}_{AB} \ - \ {{\cal T}}_{BA} \right) \ ,
\end{eqnarray}
which links the antisymmetric part of ${{\cal T}^{M}}_A$ to ${{\cal Y}^M}_{AB} $. In addition,
diffeomorphisms act on vielbein and spin connection according to
\begin{eqnarray}
&&  \delta\,{e_M}^A \ = \ D_M\, \xi^A \ - \ {S^A}_{MN} \, \xi^N \ , \nonumber \\  && \delta\, {\omega_M}^{AB} \ = \ - \ {R_{MN}}^{AB} \, \xi^N \ ,
\eea{}
and yield a second Bianchi identity,
\bea
D_M\,{{\cal T}^M}_N &+& {S^P}_{MN}\,{{\cal T}^M}_P \,- \, {S^P}_{PM} \,{{\cal T}^M}_{N}  \nonumber \\  &=& - \ {R_{MN}}^{AB}\ {{\cal Y}^M}_{AB} \ ,
\eea
which links the Riemann tensor and ${{\cal Y}^M}_{AB} $ to the covariant divergence of the energy--momentum tensor.
Notice that the torsion tensor accompanies all covariant divergences present in these equations. This is due to a subtlety related to total derivatives, whose link to covariant divergences is affected by torsion, and reads
\beq
D_M\,V^M \ = \ {S^{M}}_{MN}\, V^N \ + \ \frac{1}{e}\ \partial_M\left( e\, V^M \right)  \ . \label{tot_der}
\eeq

Retracing these arguments for the Einstein--Hilbert action
\beq
{\cal S}_{EH} \ = \ \frac{1}{2\,k^2}\ \int_{\cal M} d^{D}x \ e \ {e^M}_A\,{e^N}_B\,{R_{MN}}^{AB} \eeq
leads to
\beq
\delta\,{\cal S}_{EH}  \ = \  - \ \frac{1}{k^2} \int_{\cal M} d^{D}x \ e \left[ \delta\,{\omega_N}^{AB} \,{\Theta^N}_{AB} \ + \ \delta\,{e_M}^A\,{G^M}_A \right] \ ,
\eeq
where
\beq
{G^M}_A  = \left({e^M}_C\,{e^P}_A \, - \, \frac{1}{2}\,{e^M}_A\,{e^P}_C \right) {e^Q}_D\, {R^{CD}}_{PQ}
\eeq
is a \emph{generally non--symmetric} Einstein tensor, and
\beq
{\Theta^N}_{AB} \ = \ - \ \frac{1}{2}\left({S^P}_{PA} \ {e^N}_B \ - \ {S^P}_{PB} \ {e^N}_A \right) \ - \ \frac{1}{2}\ {S^N}_{AB} \ . \label{theta_s}
\eeq
The counterparts of the preceding Bianchi identities,
\bea
D_M\, {\Theta^M}_{AB} &-& {S^P}_{PM} \,{\Theta^M}_{AB} \ = \ \frac{1}{2} \left( {G}_{AB} \ - \ {G}_{BA} \right)  \ , \nonumber \\
D_M\,{G^M}_N &+& {S^P}_{MN}\,{G^M}_P\ - \ {S^P}_{PM} \,{G^M}_{N}  \nonumber \\  &=& - \ {\Theta^M}_{AB}\,{R_{MN}}^{AB} \ , \label{bianchi_grav}
\eea
guarantee the consistency of the Einstein equations
\beq
{G^M}_A \ = \ 2\, k^2\, {{\cal T}^M}_A \ , \qquad
{\Theta^M}_{AB} \ = \ 2\, k^2\,{{\cal Y}^M}_{AB} \ .
\eeq

We can now elaborate on how the well--known link between Killing vectors, energy--momentum tensor and conserved currents is affected by the presence of torsion. Retracing familiar steps leads to define the Killing vectors $\zeta_M$ and a new tensor,
$\theta^{AB}$, and global symmetries are identified demanding that their contributions conspire to leave both the vielbein and the spin connection unaffected:
\bea
\delta\,{e_M}^A &\equiv& \ D_M\, \zeta^A \ - \ {S^A}_{MN} \, \zeta^N \ + \ \theta^{AB}\, e_{NB} \ = \ 0 \ , \nonumber \\
\delta\, {\omega_M}^{AB} &\equiv& - \ {R_{MN}}^{AB} \, \zeta^N \ - \ D_M\,\theta^{AB} \ = \ 0 \ . \label{deltaeomega}
\eea
The first condition can be turned into
\beq
\theta^{AB} \ = \ D^A\, \zeta^B \ - \ {S^{BA}}_C\, \zeta^C \ , \label{epsilonAB}
\eeq
while the antisymmetry of $\theta^{AB}$ translates into the \emph{modified Killing equation}
\beq
D_M\,\zeta_N \ + \ D_N\,\zeta_M \ = \ \left( {S_{MN}}^P \ + \ {S_{NM}}^P\right)\zeta_P  \ . \label{mod_killing}
\eeq
Notice also that, using eq.~\eqref{epsilonAB}, the second of eqs.~\eqref{deltaeomega} can be cast in the form
\beq
D_M\,D_A\,\zeta_B \ = \ \left(D_M\,{S_{BA}}^N\right) \zeta_N \ + \ {S_{BA}}^N\,D_M\,\zeta_N \ - \ R_{MNAB}\,\zeta^N \ ,
\eeq
which generalizes the familiar result for the second derivatives of Killing vectors.

Noether currents should now satisfy the \emph{modified conservation laws}
\beq
D_M\,{\cal J}^M \ - \ {S^M}_{MN}\, {\cal J}^N \ = \ 0 \ , \label{mod_conserv}
\eeq
a subtlety whose origin we already highlighted in eq.~\eqref{tot_der}.  Given a Killing vector $\zeta^A$ solving eq.~\eqref{mod_killing}, one can verify that the combinations
\beq
{\cal J}^M \ = \ {{\cal T}^M}_N\, \zeta^N \ - \ {{\cal Y}^M}_{AB}\, \theta^{AB} \ , \label{noether}
\eeq
with $\theta^{AB}$ given by eq.~\eqref{epsilonAB}, are the Noether currents we are after, and satisfy eq.~\eqref{noether}.

When ${\cal M}$ has a boundary $\partial {\cal M}$, the time dependence of a Noether charge depends crucially on the boundary conditions, since
\beq
\frac{d\,Q(t)}{dt} \ = \ \int_{\partial{\cal M}}  d^{D-1} x \, \delta(x^0 - t)\,\sqrt{-g} \ {\cal J}^r \ ,
\eeq
and therefore the condition
\beq
\sqrt{-g}\,\left. {\cal J}^r \right|_{\partial {\cal M}} \  = \ 0 \ , \label{bc_bose}
\eeq
is needed to prevent the charge from flowing across the boundary. This crucial condition underlies the analysis sketched in the preceding section, and translates into the independent sets of conditions
\beq{}
 \sqrt{-g}\, \left. {{\cal T}^r}_N \, \zeta^N\right|_{\partial {\cal M}} \ = \ 0 \, , \quad \sqrt{-g}\, \left. {{\cal Y}^r}_{AB} \,\theta^{AB} \right|_{\partial {\cal M}} \ = \ 0 \ . \label{bcs}
\eeq
Notice that, in contrast with what we did in~\cite{ms_20}, here we kept the factor $\sqrt{-g}$ in the final form of these conditions. This factor plays indeed an important role in the examples at stake, due to the singularities present at the ends of the $r$-interval.

In the backgrounds of interest, which rest on the class of metrics in eq.~\eqref{metric_space}, one is to insist on translational symmetries in spacetime and in the internal torus, which are granted for Bose fields by the two sets of conditions
\bea{}
&& \sqrt{-g}\, \left. {{\cal T}^r}_\mu \right|_{\partial {\cal M}} \ = \ 0 \ , \quad
 \sqrt{-g}\, \left. {{\cal T}^r}_i \right|_{\partial {\cal M}} \ = \ 0 \ ,
\eea
and on Lorentz symmetries in spacetime.
For the bosonic perturbations of the background of eqs.~\eqref{4d_inter_flux_H2}, once an ${\cal A}^\dagger\,{\cal A}$ form is reached within a sector of the spectrum, these conditions translate into the demand that
\beq{}
\left. \Phi \,{\cal A}\,\Phi \right|_{\partial {\cal M}} \ = \ 0 \ ,
\eeq
and boundary conditions of this type grant the positivity of the mass spectrum.

Fermi fields require an additional effort, aimed at enforcing the second of eqs.~\eqref{bcs}. This grants the Lorentz symmetry in the resulting Minkowski spacetime. One can do it via a matrix $\Lambda$ subject to the conditions,
\beq
\Lambda^2 \ = \ 1\ , \qquad \left\{ \Lambda \,,\, \gamma^0\gamma^r\right\} \ = \ 0 \label{Lambda_constr_kin}
\eeq
which eliminate the contributions arising from varying Dirac--like actions, together with
\beq
\left[ \Lambda , \gamma_{\mu\nu} \right] \ = \ 0 \  , \label{constraints_Lambda}
\eeq
which grant indeed the conservation of the Lorentz charges, once the preceding conditions hold. However, when the spinors are subject to Weyl, Majorana, or Majorana--Weyl conditions, there are further descriptions. Thus, intriguingly, there is no assignment granting the preservation of the full set of Lorentz
charges in the Dudas--Mourad setting beyond six dimensions, due to the Majorana--Weyl nature of the ten--dimensional spinors present in the three ten--dimensional models at stake. Our analysis of the preceding section involves the type--$IIB$ theory, whose two Majorana--Weyl gravitinos can be mixed by
\beq{}
\Lambda \ = \ i \,\gamma^0\,\gamma^1\,\gamma^2\,\gamma^3 \ \sigma_2 \ ,
\eeq{}
with $\sigma_2$ the Pauli matrix, and all translational symmetries can be preserved, together with the Lorentz symmetries of the resulting four--dimensional spacetime, once these conditions are enforced.

\section{Conclusions}

The three tachyon--free ten--dimensional string models with broken supersymmetry have already provided some interesting lessons. Their main lessons so far, in our opinion, are the following:
\begin{itemize}
    \item the cosmological solutions driven by the tadpole potential grant a weak--coupling phase after the initial singularity that has some features suggestive of an intriguing mechanism to inject an inflationary phase;
    \item while symmetric vacua with broken supersymmetry are fraught with instabilities, there is some room for stable non--symmetric compactifications, above and beyond what was found in~\cite{bms} for Dudas--Mourad vacua, since there is some freedom to adjust their internal scales. It will be interesting to explore further these directions;
    \item we have managed to exhibit scenarios where supersymmetry is broken, the internal $r$-direction is compact and the lower--dimensional Planck mass is finite, while the string coupling has an upper bound. This result, however, was obtained within the type--$IIB$ theory, and we have gathered some clear evidence that the string coupling is not bounded in the presence of the tadpole potentials~\eqref{tadpole_potential}: asymptotically, they lead to what was seen in the original Dudas--Mourad vacuum or, as stressed in~\cite{ivano}, to the ``closure'' of spacetime.
    \item In contrast with these encouraging results, we have not managed to eliminate curvature singularities, in any of the settings that we have explored. A natural expectation is that this might be possible taking into account systematically the curvature corrections of String Theory. However, the preliminary analysis in~\cite{dc} is not encouraging in this respect.
\end{itemize}
\begin{figure}[ht]
\includegraphics[width=0.25\textwidth]{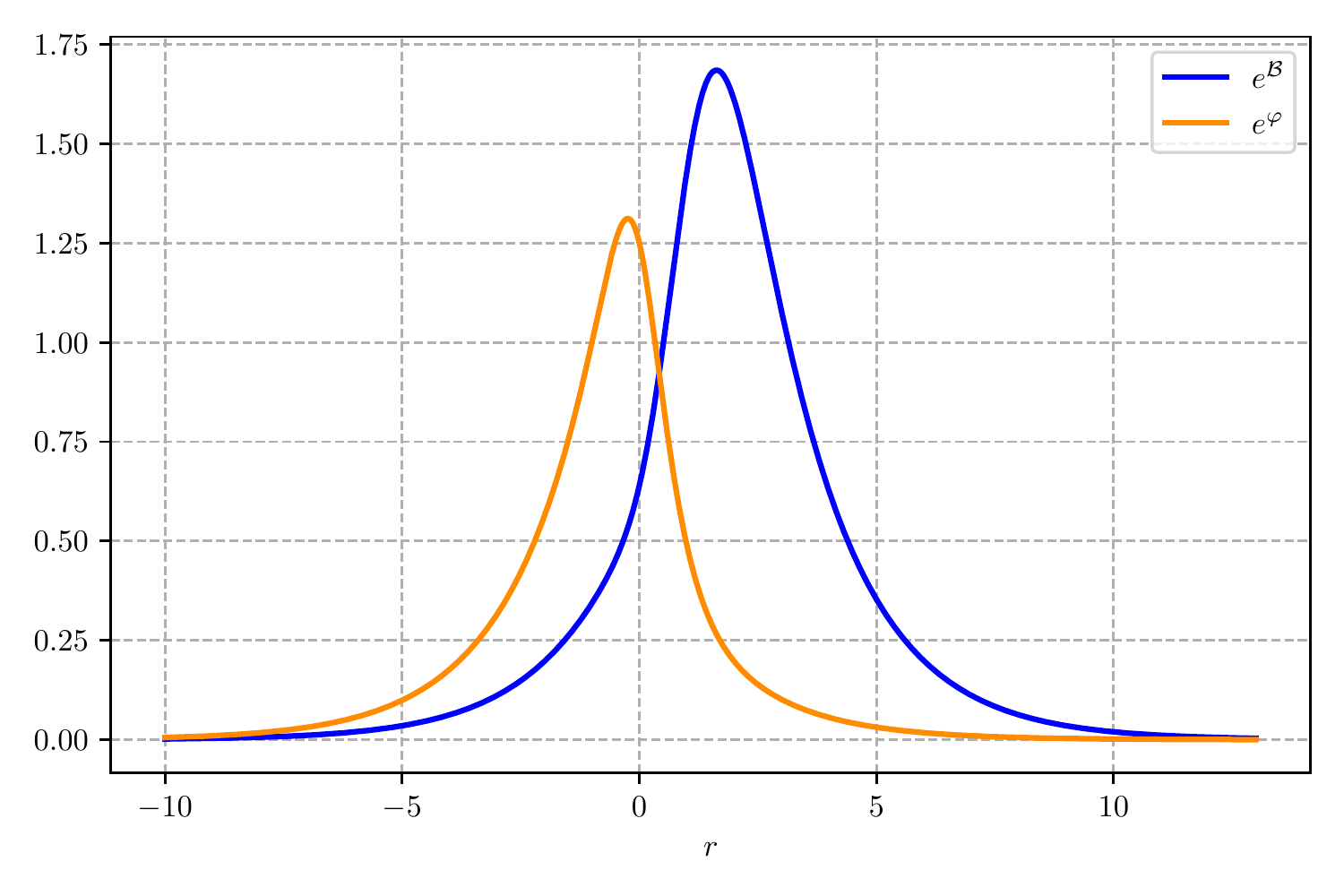}
\centering
\caption{ \small The two functions $e^{\cB}$ ( equal to $e^{\cA}$ for this model) and $e^{\frac{4}{3}\,\phi}$ for the potential of eq.~\eqref{dm_integrable2}. In this example the internal length is finite and the coupling $g_s$ is bounded, but the curvature remains singular.}
\label{fig:fourth_inverted}
\end{figure}

Let us conclude by describing what is perhaps a skew attempt~\cite{ps} at bypassing the strong--coupling problem, in the belief that it might be solved by suitable quantum corrections. The issue is: which types of modifications of the potential of eq.~\eqref{tadpole_potential} could grant this result?
Relying on a beautiful set of integrable dynamical systems already explored in connection with the climbing mechanism~\cite{fss}, we have explored toy compactifications in the presence of a wide family of potentials, identifying the rationale behind those that do grant the desired property. The strong--coupling problem can indeed disappear, provided the potential is \emph{steep enough} and \emph{unbounded from below}! For example, the ``corrected'' potential
\beq{}
V \ = \ V_0\left(e^{\frac{3}{4}\,\phi} \ - \ e^{3\,\phi} \right) \label{dm_integrable2}
\eeq{}
yields the Dudas--Mourad--like vacuum solutions
\bea
	e^{\mathcal A} &=&  e^{\mathcal A_0} \frac{\big[ \cosh(\omega (r - r_{\hat \vf})) \big]^{\frac{\gamma^2}{1 - \gamma^2}}}{\big[ \cosh(\gamma \omega (r - r_{\hat{\mathcal A}}))  \big]^{\frac{1}{1 - \gamma^2}}} \ , \nonumber \\
 \qquad e^{\frac{4}{3}\,\vf} &=&  e^{\vf_0} \frac{\big[ \cosh(\gamma \omega (r - r_{\hat{\mathcal A}}))  \big]^{\frac{\gamma}{1 - \gamma^2}}}{\big[ \cosh(\omega (r - r_{\hat \vf}))  \big]^{\frac{\gamma}{1 - \gamma^2}}} \ . \label{fourth_cosh}
\eea
that are depicted in fig.~\ref{fig:fourth_inverted}.
\begin{figure}[ht]
\includegraphics[width=0.25\textwidth]{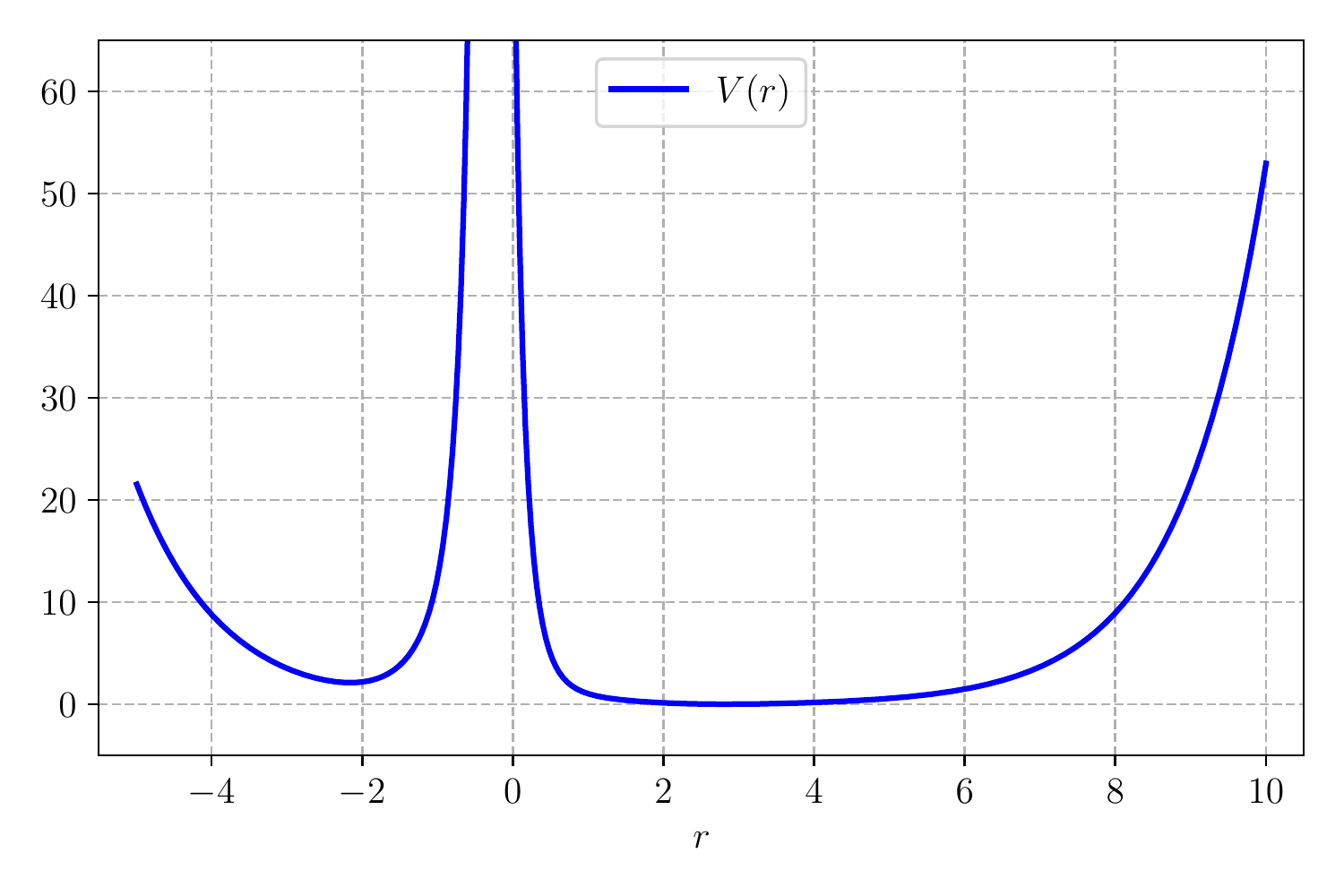}
\centering
\caption{ \small Surprisingly, the potential $b$ of eq.~\eqref{AAdagger} that emerges from the solution of fig.~\ref{fig:fourth_inverted} is always \emph{positive}, despite its origin from a potential $V(\phi)$ that is unbounded from below. Consequently, the eigenvalues of the corresponding Schr\"odinger problem, which are to lie above its minimum, are also positive, and the scalar perturbations for this model system contain no tachyonic modes.}
\label{fig:stability_fourth}
\end{figure}

Retracing the steps that led to eq.~\eqref{AAdagger}, and despite the presence of a negatively unbounded potential to begin with, the study of scalar perturbations in this example can be recast in the form of eq.~\eqref{AAdagger}, with the \emph{positive} $b$ function depicted in fig.~\ref{fig:stability_fourth}, so that the model is perturbatively stable. The mechanism that underlies these results is actually the Euclidean counterpart of the climbing mechanism in the presence of steep positive exponential potentials reviewed in Section~\ref{ref:cosmo}: potentials are indeed inverted when moving to Euclidean settings, and this is precisely what one does in the analytic continuation that we effected at the beginning to connect eqs.~\eqref{dm_cosmo} and \eqref{dm_spatial}.

\begin{figure}[ht]
\centering
\includegraphics[width=40mm]{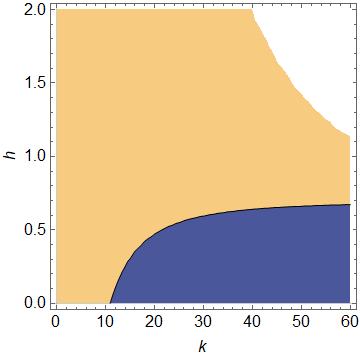}
\caption{\small A typical region of instability that emerged from two--parameter tests for singlet scalar perturbations with $\mathbf{k} \neq 0$, bounded from above by a value of $h$ of order one. Within the blue region the system has complex eigenvalues, and therefore unstable modes, for all allowed values of $\mathbf{k} \neq 0$, which are treated here, for simplicity, as a continuum.}
\label{fig:region}
\end{figure}

One may wonder whether the type of Kaluza--Klein instabilities that we have discussed are the only difficulty that one may be confronted with. In particular, are Hermitian Schr\"odinger-type systems the inevitable endpoint of these investigations? The answer is no, as will be explained in detail in~\cite{ms21_2}. The spectrum of singlet scalar perturbations with $\mathbf{k}\neq 0$ in the fluxed compactifications addressed in Section~\ref{sec:static} contains indeed two novel features: it depends on a dimensionless parameter, $h=\frac{\left|H_5\right| \rho}{\sqrt{2}}$, which sets the scale of the five--form field strength, and the resulting Schr\"odinger potential is a \emph{non--symmetric} matrix. As a result, complex eigenvalues are generically present, although we have found some evidence, in variational tests, that they disappear for $h> h_{min}$, with $h_{min}$ a value of order one (see fig.~\ref{fig:region}). Complex eigenvalues signal instabilities, and actually of a far worse kind than those exhibited so far, since their tachyonic masses grow, in absolute value, with the Kaluza--Klein momentum $\mathbf{k}$! Still, for $h>h_{min}$ these instabilities disappear, and the system passes all perturbative tests that we have been able to perform to date, provided that, as we have sketched in Section~\ref{sec:static}, the scale of the internal torus is small enough compared to the scale $\rho$ that enters eqs.~\eqref{4d_inter_flux_H2} and is proportional to the length $\ell=\rho\,h^\frac{1}{4}$  of interval. These conditions translate into the inequality $\frac{R}{\rho}\,<\,\eta_c$ where $\eta_c$ is of order $10^{-2}$, and these effective field theory considerations are thus reliable provided, say, $\rho \sim \left(10^4 - 10^5\right) \sqrt{\alpha'}$, which leaves an ample window for them.

In conclusion, we have collected some evidence that non--symmetric internal spaces may allow potentially interesting stable string compactifications to Minkowski space with broken supersymmetry. One cannot fail to notice that the lack of internal symmetries, together with the flatness of the internal tori, are reminiscent of the Calabi--Yau setup that first linked the supersymmetric hexagon of fig.~\ref{fig:susyduality} to four dimensions. The path is still long and fraught with potential pitfalls, curvature singularities appear inevitable and strong--coupling regions surface here and there, but these investigations provide, in our opinion, some encouragement for the prospects of broken supersymmetry in String Theory.
Time and more work will clarify whether or not this mild optimism is well grounded.

\vskip 24pt
\section*{Acknowledgments}

\vskip 12pt

We are grateful to E.~Dudas for stimulating discussions and a long and fruitful collaboration, to I.~Basile for a collaboration that led to~\cite{bms} and to P.~Pelliconi for a collaboration that led to~\cite{ps}. In addition, we would like to thank G.~Bogna and Y.~Tatsuta, who are investigating with us related aspects of the problem, and S.~Raucci, who is investigating with us other aspects of broken supersymmetry and pointed out several misprints in an earlier version of the manuscript. The work of AS was supported in part by Scuola Normale, by INFN (IS GSS-Pi) and by the MIUR-PRIN contract 2017CC72MK\_003. JM is grateful to Scuola Normale Superiore for the kind hospitality extended to him while work reviewed here was in progress. AS is grateful to APC and DESY--Hamburg for the kind hospitality, and to the Alexander von Humboldt foundation for the kind and generous support, while work reviewed here was in progress.

\end{document}